\documentclass[twocolumn,times]{aastex631}

\usepackage{graphicx}
\usepackage{amsfonts,amsmath,amssymb}
\usepackage{xspace}
\usepackage{bm}
\usepackage[normalem]{ulem}
\usepackage{color,soul}

\let\tablenum\relax  % fix conflict with aastex
\usepackage{siunitx}

\DeclareSIUnit{\erg}{erg}
\DeclareSIUnit{\msun}{M_\odot}
\DeclareSIUnit{\density}{\gram\per\cm\cubed}
\DeclareSIUnit{\luminosity}{\erg\per\second}

\newcommand{\tenx}{10$\times{}$}
\newcommand{\onex}{0.01$\times{}$}

\newcommand{\snowglobes}{\textsc{SNOwGLoBES}}
\newcommand{\nulib}{\textsc{NuLib}}
\newcommand{\flash}{\textsc{FLASH}}
\newcommand{\stir}{\textsc{STIR}}
\newcommand{\flashbang}{\textsc{flashbang}}
\newcommand{\flashsnowglobes}{\textsc{flash\_snowglobes}}

\newcommand{\argon}{\ensuremath{\mathrm{^{40}Ar}}}
\newcommand{\krypton}{\ensuremath{\mathrm{^{86}Kr}}}
\newcommand{\chlorine}{\ensuremath{\mathrm{^{40}Cl}}}
\newcommand{\potassium}{\ensuremath{\mathrm{^{40}K}}}
\newcommand{\helium}{\ensuremath{\mathrm{^{3}He}}}

\newcommand{\alphL}{\ensuremath{\alpha_\Lambda}}
\newcommand{\Ye}{\ensuremath{Y_e}}
\newcommand{\Yl}{\ensuremath{Y_l}}
\newcommand{\MPNS}{\ensuremath{M_\mathrm{PNS}}}
\newcommand{\Mcore}{\ensuremath{M_\mathrm{core}}}
\newcommand{\Mfe}{\ensuremath{M_\mathrm{Fe}}}
\newcommand{\rshock}{\ensuremath{r_\mathrm{sh}}}
\newcommand{\tBH}{\ensuremath{t_\mathrm{BH}}}
\newcommand{\velx}{\ensuremath{v_r}}
\newcommand{\etaheat}{\ensuremath{\eta_\mathrm{heat}}}
\newcommand{\rshmax}{\ensuremath{\mathrm{max}(\rshock)}}
\newcommand{\tbounce}{\ensuremath{t_b}}
\newcommand{\mdotbounce}{\ensuremath{\dot{M}_b}}
\newcommand{\Yebounce}{\ensuremath{Y_e}}
\newcommand{\tnue}{\ensuremath{t_{\nu_e}}}
\newcommand{\gpot}{\ensuremath{V_b}}

\newcommand{\nue}{\ensuremath{\nu_{e}}}
\newcommand{\nuebar}{\ensuremath{\bar{\nu}_e}}
\newcommand{\nux}{\ensuremath{\nu_x}}
\newcommand{\Lnu}{\ensuremath{L_\nu}}
\newcommand{\Lnue}{\ensuremath{L_{\nue}}}

\newcommand{\Lnux}{\ensuremath{L_{\nux}}}
\newcommand{\Enu}{\ensuremath{{\langle E_{\nu} \rangle}}}
\newcommand{\Enue}{\ensuremath{{\langle E_{\nue} \rangle}}}
\newcommand{\Enux}{\ensuremath{{\langle E_{\nux} \rangle}}}
\newcommand{\nuEtot}{\ensuremath{{\langle E \rangle}}}
\newcommand{\elec}{\ensuremath{e^-}}
\newcommand{\pos}{\ensuremath{e^+}}
\newcommand{\nuecc}{\ensuremath{\nue \mathrm{CC}}}
\newcommand{\nuebcc}{\ensuremath{\nuebar \mathrm{CC}}}
\newcommand{\nushock}{\ensuremath{t_{\nue} - \tbounce{}}}
\newcommand{\Fe}{\ensuremath{F_e}}
\newcommand{\Febar}{\ensuremath{\bar{F}_e}}
\newcommand{\Fx}{\ensuremath{F_x}}
\newcommand{\Fxbar}{\ensuremath{\bar{F}_x}}

\newcommand*{\PAMSU}{Department of Physics and Astronomy, Michigan State University, East Lansing, MI 48824, USA}
\newcommand*{\CMSE}{Department of Computational Mathematics, Science, and Engineering, Michigan State University, East Lansing, MI 48824, USA}
\newcommand*{\NSCL}{National Superconducting Cyclotron Laboratory, Michigan State University, East Lansing, MI 48824, USA}
\newcommand*{\JINA}{Joint Institute for Nuclear Astrophysics -- Center for the Evolution of the Elements, Michigan State University, East Lansing, MI 48824, USA}
\newcommand*{\NCSU}{Department of Physics, North Carolina State University, Raleigh, NC 27695, USA}
\newcommand*{\OKSU}{The Oskar Klein Centre, Department of Astronomy, Stockholm University, AlbaNova, SE-106 91 Stockholm, Sweden}

\received{18 February 2022}
\revised{9 September 2022}
\accepted{16 September 2022}
\submitjournal{ApJ}

\shorttitle{Comparison of EC Rates in CCSN Simulations}
\shortauthors{Johnston et al.}

\begin{document}

\title{Comparison of Electron Capture Rates in the N=50 Region using 1D Simulations of Core-collapse Supernovae}

\author[0000-0003-4023-4488]{Zac Johnston}
\affiliation{\PAMSU} 
\affiliation{\JINA}
\email{zacjohn@msu.edu}

\author[0000-0001-9213-0117]{Sheldon Wasik}
\affiliation{\PAMSU}

\author{Rachel Titus}
\affiliation{\PAMSU} 
\affiliation{\NSCL}

\author[0000-0001-9440-6017]{MacKenzie L. Warren}
\affiliation{\PAMSU} 
\affiliation{\JINA}
\affiliation{\NCSU}

\author[0000-0002-8228-796X]{Evan P. O'Connor}
\affiliation{\OKSU}

\author[0000-0001-6076-5898]{Remco Zegers}
\affiliation{\PAMSU}
\affiliation{\JINA}
\affiliation{\NSCL} 

\author[0000-0002-5080-5996]{Sean M. Couch}
\affiliation{\PAMSU} 
\affiliation{\JINA}
\affiliation{\NSCL} 
\affiliation{\CMSE} 

%==============================================================
%         Abstract
%==============================================================
\begin{abstract}
  Recent studies have highlighted the sensitivity of core-collapse supernovae (CCSNe) models to electron-capture (EC) rates on neutron-rich nuclei near the $N=50$ closed-shell region.
  In this work, we perform a large suite of one-dimensional CCSN simulations for 200 stellar progenitors using recently updated EC rates in this region. 
  For comparison, we repeat the simulations using two previous implementations of EC rates: a microphysical library with parametrized $N=50$ rates (LMP), and an older independent-particle approximation (IPA).
  We follow the simulations through shock revival up to several seconds post-bounce, and show that the EC rates produce a consistent imprint on CCSN properties, often surpassing the role of the progenitor itself.
  Notable impacts include the timescale of core collapse, the electron fraction and mass of the inner core at bounce, the accretion rate through the shock, the success or failure of revival, and the properties of the central compact remnant.
  We also compare the observable neutrino signal of the neutronization burst in a DUNE-like detector, and find consistent impacts on the counts and mean energies.
  Overall, the updated rates result in properties that are intermediate between LMP and IPA, and yet slightly more favorable to explosion than both.
\end{abstract}

%==============================================================
%         Introduction
%==============================================================
\section{Introduction}
\label{s.intro}
% background
Massive stars ($\gtrsim \SI{8}{\msun}$) are destined to undergo iron core-collapse, either imploding entirely into a black hole (BH), or violently ejecting their outer layers and leaving behind a proto-neutron star (PNS) in a core-collapse supernova \citep[CCSN; see reviews in][]{janka:2007,muller:2020}.
Of the myriad physical processes that contribute to these stellar deaths, the capture of electrons onto protons via the weak interaction plays a central role.

Electron-capture (EC) regulates the deleptonization of nuclear matter during collapse, and thus helps to set the initial conditions of the shock at core bounce \citep[see reviews in][]{langanke:2003a,langanke:2021}.
The uncertainties on electron capture rates can span orders of magnitude and produce larger variations in core-collapse properties than changes to the nuclear equation of state (EOS) or stellar progenitor \citep{sullivan:2016, pascal:2020}.

% EC rates, CCSN studies
It is experimentally and computationally difficult to constrain EC rates under astrophysical conditions, especially for the heavy, neutron-rich nuclei relevant to core-collapse.
For this reason, CCSN simulations typically rely on parametrized approximations, particularly those of \citet{bruenn:1985} and \citet{langanke:2003}.
Recent decades, however, have seen the continued development of tabulated rates for larger numbers of nuclei based on shell-model calculations \citep[e.g.,][]{oda:1994,langanke:2000, langanke:2003, suzuki:2016}.

%% N=50 sensitivity
A systematic study by \citet{sullivan:2016} showed that core-collapse simulations were most sensitive to changes in the electron capture rates of neutron-rich nuclei near the $N = 50$, $Z=28$ closed-shell region.
The drawback is that most of these rates relied on the parametrization of \citet{langanke:2003}, which is extrapolated from rates on nuclei near the valley of stability.
In a follow-up study focusing on 74 nuclei in the high-sensitivity region, \citet{titus:2018} showed that these rates are likely overestimated by up to two orders of magnitude, further emphasizing the need for updated rates. 
Using new experimental measurements of the $\krypton (t,\helium+\gamma)$ charge-exchange reaction, \citet{titus:2019} calculated microphysical rates on 78 nuclei in this region, confirming that the parametrized rates are overestimated.

In the meantime, \citet{raduta:2017} improved upon the parametrization of \citet{langanke:2003} for neutron-rich nuclei by accounting for temperature, electron density, and odd-even effects.
Comparing one-dimensional (1D) core-collapse simulations using this improved parametrization, \citet{pascal:2020} demonstrated the expected increase in core mass and electron fraction due to an average decrease in EC rates.
\citet{pascal:2020} also performed a rate sensitivity study, independently verifying the findings of \citet{sullivan:2016} and \citet{titus:2018} that EC rates in the $N=50$ region remain the most crucial for CCSNe.

% paper summary
In this paper, we present simulations of CCSNe through to shock revival/failure for 200 stellar progenitors using the updated $N=50$ rates from \citet{titus:2019}.
We run corresponding simulations using the baseline rate library from \citet{sullivan:2016} with the improved approximation of \citet{raduta:2017}, and a third set using the independent-particle approximation (IPA) of \citet{bruenn:1985}.
By comparing the model sets, we investigate the impact of updated EC rates on core collapse, shock revival, and neutrino emission across a variety of progenitors.

% structure of paper
The paper is structured as follows.
In Section~\ref{s.methods} we describe our methods, including the EC rate tables (\S~\ref{ss.methods.ecrates}), the setup of the CCSN simulations (\S~\ref{ss.methods.flash}), and the calculation of observable neutrino signals (\S~\ref{ss.methods.snowglobes}).
In Section~\ref{s.results}, we compare the simulation results, with a detailed comparison of three reference progenitors (\S~\ref{ss.results.reference}), the impact across the full progenitor population (\S~\ref{ss.results.population}), the compact remnants (\S~\ref{ss.results.remnants}), and the predicted neutrino signal (\S~\ref{ss.results.snowglobes}).
In Section~\ref{s.discussion} we interpret our results and compare them to previous studies, and give concluding remarks in Section~\ref{s.conclusion}.

%+++++++++++++++++++++++++++++++
%   (figure) Shock Radius
%+++++++++++++++++++++++++++++++
\begin{figure*}[t!]
  \centering
  \includegraphics[width=\linewidth]{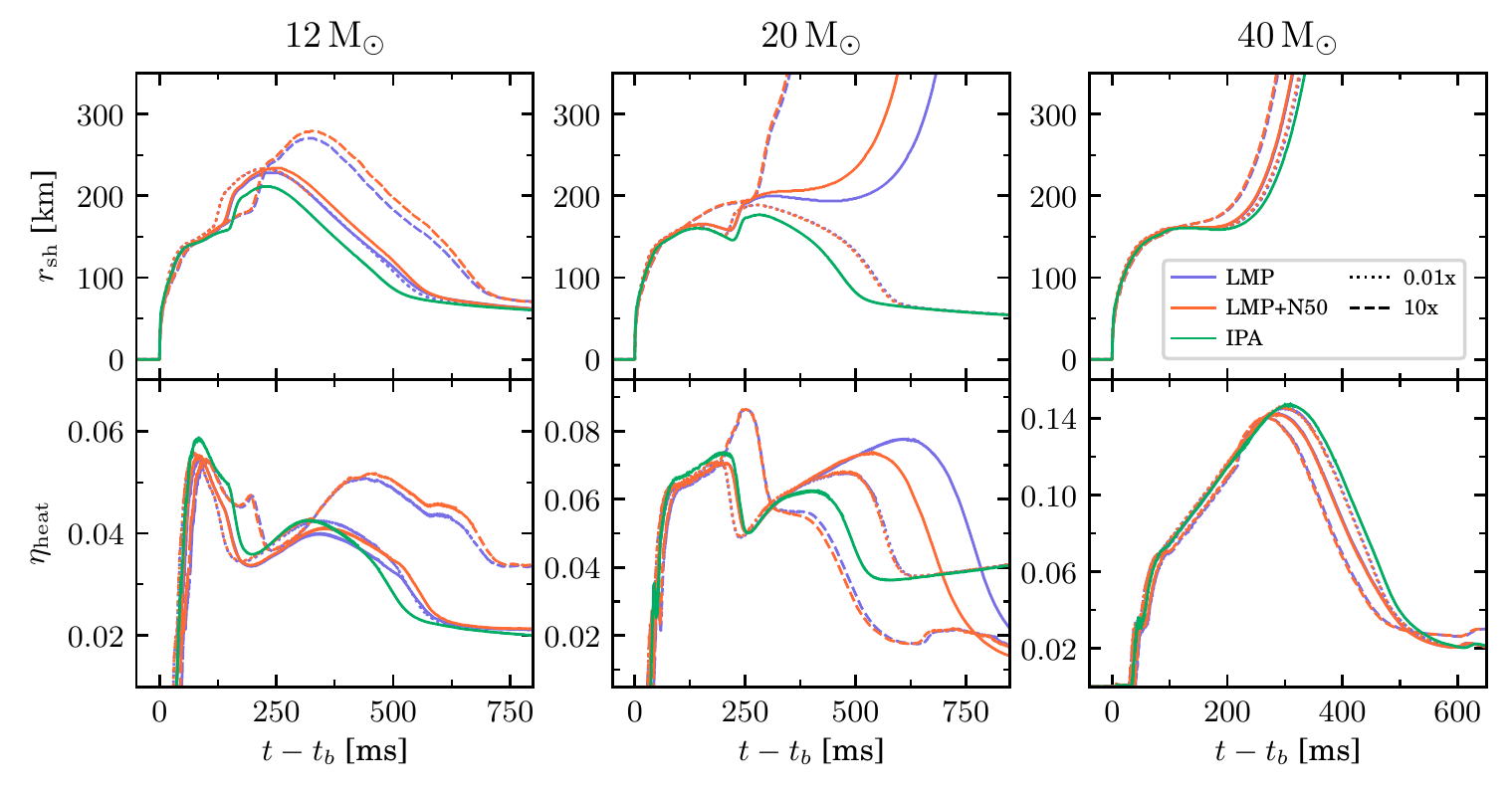}
  \caption{Post-bounce evolution of shock radius (\rshock{}) and neutrino heating efficiency in the gain region (\etaheat{}) for three reference progenitors (\S~\ref{ss.results.reference}).
  The \onex{} and \tenx{} models consist of the LMP and LMP+N50 rates systematically scaled by factors of $0.01$ and $10$ (\S~\ref{ss.methods.ecrates}).
  Note the different \etaheat{} ranges.
  }
  \label{fig:shock_radius}
\end{figure*}

%==============================================================
%         Methods
%==============================================================
\section{Methods}
\label{s.methods}
To explore the impact of EC rates on CCSNe, we run multiple large sets of 1D simulations using the \flash{} code \citep{fryxell:2000,dubey:2009}.
For initial conditions, we use 200 stellar progenitor models from \citet{sukhbold:2016} with zero-age main-sequence (ZAMS) masses between \SIrange{9}{120}{\msun} (\S~\ref{ss.methods.flash}).

For each progenitor, we run simulations using three different implementations of EC rates, which include a independent-particle approximation, microphysical calculations, and updated experimental rates (\S~\ref{ss.methods.ecrates}).
For the \SIlist{12;20;40}{\msun} progenitors, we also run simulations with the microphysical rates scaled by factors of $0.01$ and $10$.
In total, this results in 612 supernova models evolved to between \SIrange{1}{5}{s} post-bounce.

%==========================================
%         Rate Tables
%==========================================
\subsection{EC Rates}
\label{ss.methods.ecrates}
% IPA
The first of our three EC rate sets uses the IPA on a mean nucleus \citep{fuller:1982b, bruenn:1985}.
These rates were used in the comparable \flash{} simulations of \citet{couch:2020} and \citet[though see \S~\ref{ss.methods.flash} for a note on differences]{warren:2020}.
Crucially, IPA assumes that EC completely halts for nuclei with $N \geq 40$ due to Pauli blocking, thus only permitting captures on free protons at densities above $\sim \SI{e10}{\density}$, where neutron-rich nuclei dominate the composition \citep{langanke:2003}.

% LMP
The second set, which we label LMP, uses a library of microphysical rates compiled by the National Superconducting Cyclotron Laboratory (NSCL) Charge-Exchange Group~\footnote{\url{https://github.com/csullivan/weakrates}}\textsuperscript{,}\footnote{\url{https://groups.nscl.msu.edu/charge_exchange/weakrates.html}} \citep{sullivan:2016, titus:2018}.
This library includes rates from \citet{fuller:1982b}, \citet{oda:1994}, \citet{langanke:2000}, \citet{langanke:2003}, \citet{pruet:2003} and \citet{suzuki:2016}.
For nuclei that are not covered by the above calculations, the single-state parametrization from \citet[][\textit{Model 3}]{raduta:2017} is used, which extends the approximation of \citet{langanke:2003} to account for temperature, electron density, and odd-even effects.
By unblocking EC on neutron-rich nuclei, these modern calculations replace the assumption in IPA that deleptonization is dominated at high densities by EC on free protons \citep{langanke:2003}.

%  LMP+N50
The third set, which we label LMP+N50, is the same as LMP except for updated microphysical rates in the region around $N = 50$, $Z = 28$, where core-collapse is known to be highly sensitive \citep{sullivan:2016, titus:2018, pascal:2020}.
These new rates were calculated for 78 nuclei by \citet{titus:2019} using a quasi-particle random-phase approximation (QRPA) using constraints from $(t,\helium+\gamma)$ charge-exchange experiments.
The LMP set largely relies on the \citet{raduta:2017} parametrization for these rates,  and are up to two orders of magnitude higher than the LMP+N50 set.
This is due to the fact that in the LMP+N50 set Pauli-blocking effects that reduce the rates play an important role as only transitions from the ground state are considered \citep{titus:2018, titus:2019}.

Recently, \citet{dzhioev:2020} argued that Pauli unblocking at finite temperature may actually reduce or eliminate this gap.
Indeed, during the latter stages of our present study, \citet{giraud:2022} reported new finite-temperature calculations for the $N=50$ region, which resulted in rates around an order of magnitude higher than LMP+N50 for $T \lesssim \SI{10}{GK}$ and $\rho \Ye = \SI{e11}{\density}$.
This improvement brings the rates closer to the original LMP approximation, but still lower by about a factor of 5.
The calculations by \mbox{\citet{dzhioev:2020}} and \mbox{\citet{giraud:2022}} both indicate that the temperature-dependent effects significantly increase the EC rates compared to those estimated on the basis of captures on the ground state only by \mbox{\citet{titus:2019}}.
Although a future study will be required to determine the impact of the rates developed in \mbox{\citet{giraud:2022}}, we note that the LMP+N50 rates by \mbox{\citet{titus:2019}} should be regarded as lower limits, and on the basis of \mbox{\citet{giraud:2022}}, the LMP rates provide a more realistic estimate but are likely still overestimated.

% ec-rate multiplier
For the \SIlist{12;20;40}{\msun} progenitors, in addition to the three EC rate sets above, we also systematically scale the rates of LMP and LMP+N50 following the approach used in \citet{sullivan:2016}, whereby the rates of all nuclei with atomic mass numbers of $A > 4$ are scaled by factors of $0.01$ and $10$ (hereafter labeled \onex{} and \tenx{}).

%+++++++++++++++++++++++++++++++
%   (table) Neutrino channels
%+++++++++++++++++++++++++++++++
\begin{table}[t!]
  \centering
  \caption{Neutrino detection channels used in the \snowglobes{} analysis for a DUNE-like liquid argon detector}

  \begin{tabular}{lccr}
      \hline
      Channel & Reaction & Flavor & $(\%)$\\
      \hline
      \nuecc{} & $\nue + \argon \rightarrow \elec + \potassium$ & \nue & \SIrange{70}{80}{} \\
      \nuebcc & $\nuebar + \argon \rightarrow \pos + \chlorine$ & \nuebar & $\lesssim 1$ \\
      NC & $\nu + \argon \rightarrow \nu + \argon$ & \nue, \nuebar, \nux & \SIrange{10}{20}{} \\
      ES & $\nu + \elec \rightarrow \nu + \elec$ & \nue, \nuebar, \nux & $\sim 7$ \\
      \hline
  \end{tabular}

  \tablecomments{.
  The reaction channels are charged-current (CC), neutral current (NC), and electron scattering (ES).
  Also listed are the rough percentage contributions of each channel to our total counts, which vary by model and choice of flavor mixing.
  }

  \label{tab:channels}  
\end{table}

%==========================================
%         FLASH
%==========================================
\subsection{Numerical Methods}
\label{ss.methods.flash}
To simulate the collapse and explosion of massive stars, we use the \flash{} hydrodynamics code \citep{fryxell:2000,dubey:2009} with the Supernova Turbulence in Reduced-dimensionality (\stir{}) framework \citep{couch:2020, warren:2020}.

% STIR
\stir{} enhances the \textit{explodability} of 1D CCSN models by using time-dependent mixing-length theory (MLT) to approximate convective turbulence in 1D.
We use a mixing length parameter of $\alphL = 1.25$, chosen to reproduce the convective velocities of 3D simulations, which is multiplied by the pressure scale height to obtain the mixing length \citep[for details, see][]{couch:2020}. 

% spark
We use a recently implemented hydrodynamics solver \citep{couch:2020}, which uses a fifth-order finite-volume weighted essentially non-oscillatory (WENO) spatial discretization, and a method-of-lines Runge-Kutta time integration. 

% M1
For neutrino transport, we use the ``M1'' scheme, an explicit two-moment method with an analytic closure \citep[described in][]{oconnor:2018b}, with three neutrino species (\nue{}, \nuebar{}, and $\nux{}=\{ \nu_\mu,\, \nu_\tau,\, \bar{\nu}_\mu,\, \bar{\nu}_\tau \}$) and 18 logarithmically spaced energy groups between \SIrange[range-phrase=\text{ and }]{1}{300}{\MeV}.

% nulib
We generate neutrino opacity tables using the open source neutrino interaction library \nulib{}\footnote{\url{https://github.com/evanoconnor/nulib}} \citep{oconnor:2015}.
The interaction rates largely follow \citet{bruenn:1985} and \citet{burrows:2006a}, with corrections for weak magnetism and nucleon recoil from \citet{horowitz:2002}.
Separate tables are calculated using the neutrino emissivities derived from each EC rate set described in Section~\ref{ss.methods.ecrates}.
We note that our tables do not include the many-body effects and virial corrections to neutrino-nucleon scattering from \citet{horowitz:2017}.
These corrections aid explodability by enhancing neutrino heating in the gain region \citep{oconnor:2017}, and thus our simulations result in fewer explosions than the corresponding models in \citet{couch:2020} and \citet{warren:2020}.
Nevertheless, this does not impede our goal of a comparison study between the EC rates.

% EOS
We use the SFHo EOS from \citet{steiner:2013}, and assume nuclear statistical equilibrium (NSE) abundances everywhere in the domain.
For the IPA EC rates, the average nucleus from the NSE distribution is used.
Self-gravity is included using an approximate general-relativistic effective potential \citep{marek:2006,oconnor:2018b}.

% progenitors
For initial conditions, we use 200 stellar progenitor models from \citet{sukhbold:2016}, the same set used with \flash{}+\stir{} in \citet{couch:2020} and \citet{warren:2020}.
These progenitors are spherically symmetric, solar-metallicity, nonrotating, and nonmagnetic, with ZAMS masses ranging from \SIrange{9}{120}{\msun}.
The set spans core compactness values of $0 \lesssim \xi_{2.5} \lesssim 0.54$ \citep[as defined in][]{oconnor:2011} and iron core masses of $1.29 \lesssim \Mfe \lesssim \SI{1.84}{\msun}$.

% domain, AMR
The simulation domain extends from the center of the star to $r = \SI{15000}{\km}$.
The domain is divided into 15 adaptive mesh refinement blocks, each containing 16 zones.
We allow up to nine levels of mesh refinement, resulting in a zone resolution of \SI{62.5}{\km} at the coarsest level and \SI{0.244}{\km} at the finest level.
The adaptive mesh refinement results in a total of roughly $1000$ zones.

%==========================================
%         SNOWGLOBES
%==========================================
\subsection{Neutrino Observables}
\label{ss.methods.snowglobes}
Following the approach used in \citet{warren:2020}, we calculate simulated observations of the neutrino burst at core bounce using \snowglobes{}\footnote{\url{https://github.com/SNOwGLoBES/snowglobes}} \citep{scholberg:2012}, which uses the \textsc{GLoBES}\footnote{\url{www.mpi-hd.mpg.de/personalhomes/globes}} \citep{huber:2005} framework to predict event rates for a given detector material.

As input for \snowglobes{}, we calculate from our simulations the neutrino flux at Earth assuming a CCSN distance of \SI{10}{kpc} and a pinched neutrino spectrum with a Fermi-Dirac parametrization \citep{keil:2003}.
We include adiabatic neutrino flavor conversions from Mikheyev--Smirnov--Wolfenstein (MSW) matter effects \citep{dighe:2000}.
For each model, we apply three separate cases of flavor mixing: no flavor mixing, normal neutrino mass ordering, and inverted mass ordering (Appendix~\ref{a.mixing}).

We calculate detection events for a \SI{40}{kt} liquid argon detector, representing the under-construction Deep Underground Neutrino Experiment \citep[DUNE;][]{abi:2021} capable of detecting large numbers of \nue{} from nearby CCSNe \citep{kato:2017}.
Table~\ref{tab:channels} summarizes the different interaction channels: charged-current (CC) reactions on \argon{} by \nue{} and \nuebar{}; neutral-current (NC) reactions on \argon{} for all flavors; and electron scattering (ES) for all flavors.
The \nuecc{} reaction channel accounts for approximately \SIrange{70}{80}{\percent} of the total counts in our models.

We capture the neutronization burst by integrating events over \SI{100}{ms} centered on the bounce, using \SI{5}{ms} time bins and \SI{0.2}{\MeV} energy bins.
For each model and flavor-mixing case, we thus obtain the total neutrino counts and the mean detected neutrino energy, \nuEtot{}, summed over all detection channels.

%+++++++++++++++++++++++++++++++
%   (figure) Bounce Profiles
%+++++++++++++++++++++++++++++++
\begin{figure*}[t!]
  \centering
  \includegraphics[width=\linewidth]{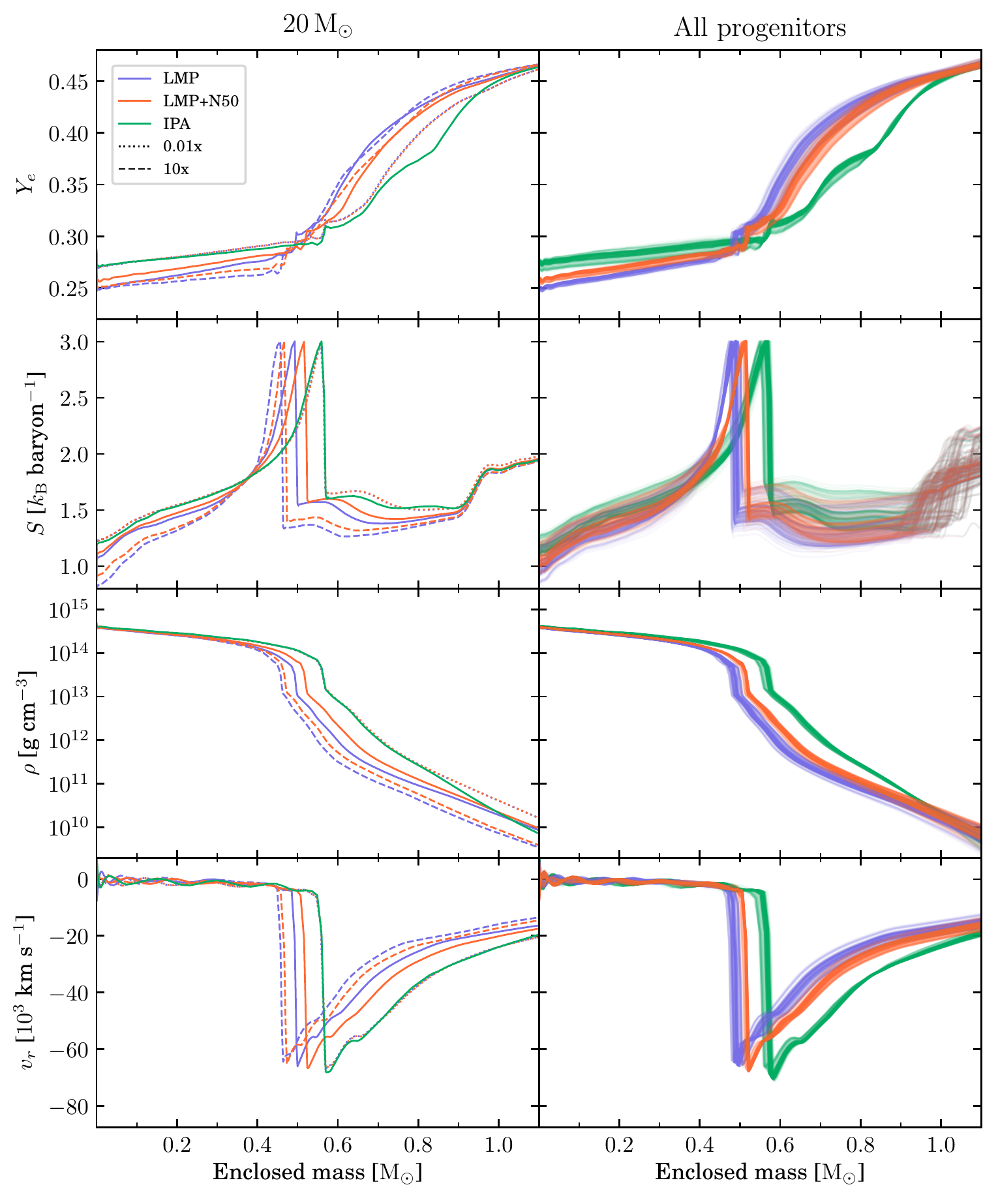}
  \caption{Radial matter profiles at core bounce versus enclosed mass for the \SI{20}{\msun} progenitor (left), and all 200 progenitors between \SIrange{9}{120}{\msun} (right).
  From top to bottom: electron fraction ($\Ye$); specific entropy ($S$); density ($\rho$); and radial velocity ($\velx$).
  Differences between the EC rate are typically larger than differences between the progenitors.
  For the \SI{20}{\msun} progenitor, the IPA and \onex{} models fail to explode, whereas the LMP, LMP+N50, and \tenx{} models successfully explode (Fig.~\ref{fig:shock_radius}).  
  }
  \label{fig:bounce_profiles}
\end{figure*}

%+++++++++++++++++++++++++++++++
%   (figure) Yl density profiles
%+++++++++++++++++++++++++++++++
\begin{figure}[t!]
  \centering
  \includegraphics[width=\linewidth]{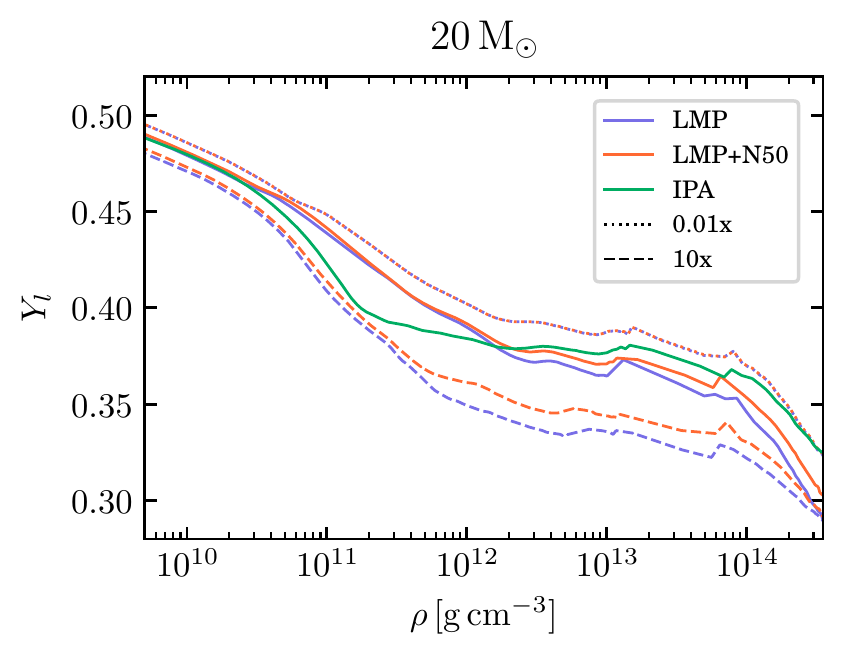}
  \caption{Lepton fraction (\Yl) versus density ($\rho$) at core bounce for the \SI{20}{\msun} progenitor.
  Of the baseline EC rate sets, IPA has the weakest deleptonization at densities $\gtrsim \SI{1e12}{\density}$, but the strongest at lower densities.
  }
  \label{fig:yl_dens}
\end{figure}

%==============================================================
%         Results
%==============================================================
\section{Results}
\label{s.results}
Our collection of 612 simulations can be sorted into three groups based on the explosion outcomes for each progenitor.
Firstly, there are those models that, for a given progenitor, fail to explode for all three EC rate sets.
Secondly, there are those with mixed explosion outcomes between the rate sets.
And thirdly, there are those that successfully explode for all three sets.

Of the 200 progenitors, 126 fail for all three rate sets, 29 have mixed explosion outcomes, and 44 explode for all sets.
The \SI{10.25}{\msun} progenitor simulations experience numerical crashes mid-shock revival, and are excluded from discussions of explosion outcome.
Of the mixed-outcome group, the IPA models always fail, whereas LMP and LMP+N50 both explode in 26 cases, and LMP+N50 is the only explosion for the remaining three cases (\SIlist{22;27.4;33}{\msun}).
In summary, there are 44 successful explosions for IPA, 70 for LMP, and 73 for LMP+N50.

The data presented here, and the codes used to analyze it, are publicly available (Appendix~\ref{a.data}).

%==========================================
%         Reference  Progenitors
%==========================================
\subsection{Reference Progenitors}
\label{ss.results.reference}
We here present detailed simulation comparisons for the \SIlist{12;20;40}{\msun} progenitors.
These progenitors are representative of the three groups of explosion outcomes, respectively: all EC rate sets fail to explode; mixed outcomes; and all successfully explode.

% r_shock
The evolution of the shock radius, \rshock{}, and the neutrino heating efficiency, $\etaheat$, are shown in Figure~\ref{fig:shock_radius}.
Here, \etaheat{} is the fraction of the total \nue{} and \nuebar{} luminosity absorbed in the gain region, which we estimate following \citet{oconnor:2011}.
All \SI{12}{\msun} models fail to explode and all \SI{40}{\msun} models successfully explode.
For the \SI{20}{\msun} models, the LMP, LMP+N50, and both \tenx{} models explode, whereas the IPA and both \onex{} models fail.

A consistent hierarchy of \rshock{} evolution is seen across the reference progenitors.
Overall, the IPA models reach the smallest \rshock{}, experience the earliest shock recession (\SIlist{12;20}{\msun}), and the latest shock revival (\SI{40}{\msun}), followed closely by the \onex{} models.
The LMP and LMP+N50 models reach larger \rshock{} before recession (\SI{12}{\msun}) and undergo earlier shock revival (\SIlist{20;40}{\msun}).
Finally, the \tenx{} models reach the largest \rshock{} before recession (\SI{12}{\msun}) and the earliest shock revivals (\SIlist{20;40}{\msun}).

For the \SI{20}{\msun} progenitor, LMP+50 appears to require a smaller heating efficiency for shock revival than LMP, suggesting more favorable explosion conditions.
The \tenx{} models experience a surge in \etaheat{} around \SI{250}{ms}, which appears to contribute to an early shock runaway.
In contrast, IPA does not reach sufficient \etaheat{} before its shock contracts, shrinking the available gain region for neutrino interactions.

% bounce profiles
The matter profiles at core bounce are shown in Figure~\ref{fig:bounce_profiles} for the \SI{20}{\msun} progenitor (left) and the full set of 200 progenitors (right).
We define bounce as the moment when the peak entropy in the core reaches $3\, k_\mathrm{B}\, \mathrm{baryon^{-1}}$.
We also define the inner core mass at bounce, \Mcore{}, as the mass enclosed within this point (also known as the homologous core mass).

As with \rshock{}, the models maintain a consistent hierarchy, even across the entire population of progenitors.
The IPA models have the largest inner core mass and electron fraction, entropy, density, and infall velocity.
This trend is followed, in order, by the \onex{}, LMP+N50, LMP, and finally \tenx{} models.
This ordering is reversed for the \Ye{} outside the shock, where IPA is the lowest and \tenx{} the largest.

% lepton fraction
In Figure~\ref{fig:yl_dens}, the lepton fraction is shown versus density at bounce for the \SI{20}{\msun} progenitor.
Above the neutrino-trapping densities of $\sim \SI{1e12}{\density}$, the \onex{} models have the largest \Yl{}, followed by IPA, LMP+N50, LMP, and \tenx{}.

%++++++++++++++++++++++++++++++++++++++++++++++++++++
%   (figure) pop stats
%++++++++++++++++++++++++++++++++++++++++++++++++++++
\begin{figure}[t!]
  \centering
  \includegraphics[width=\linewidth]{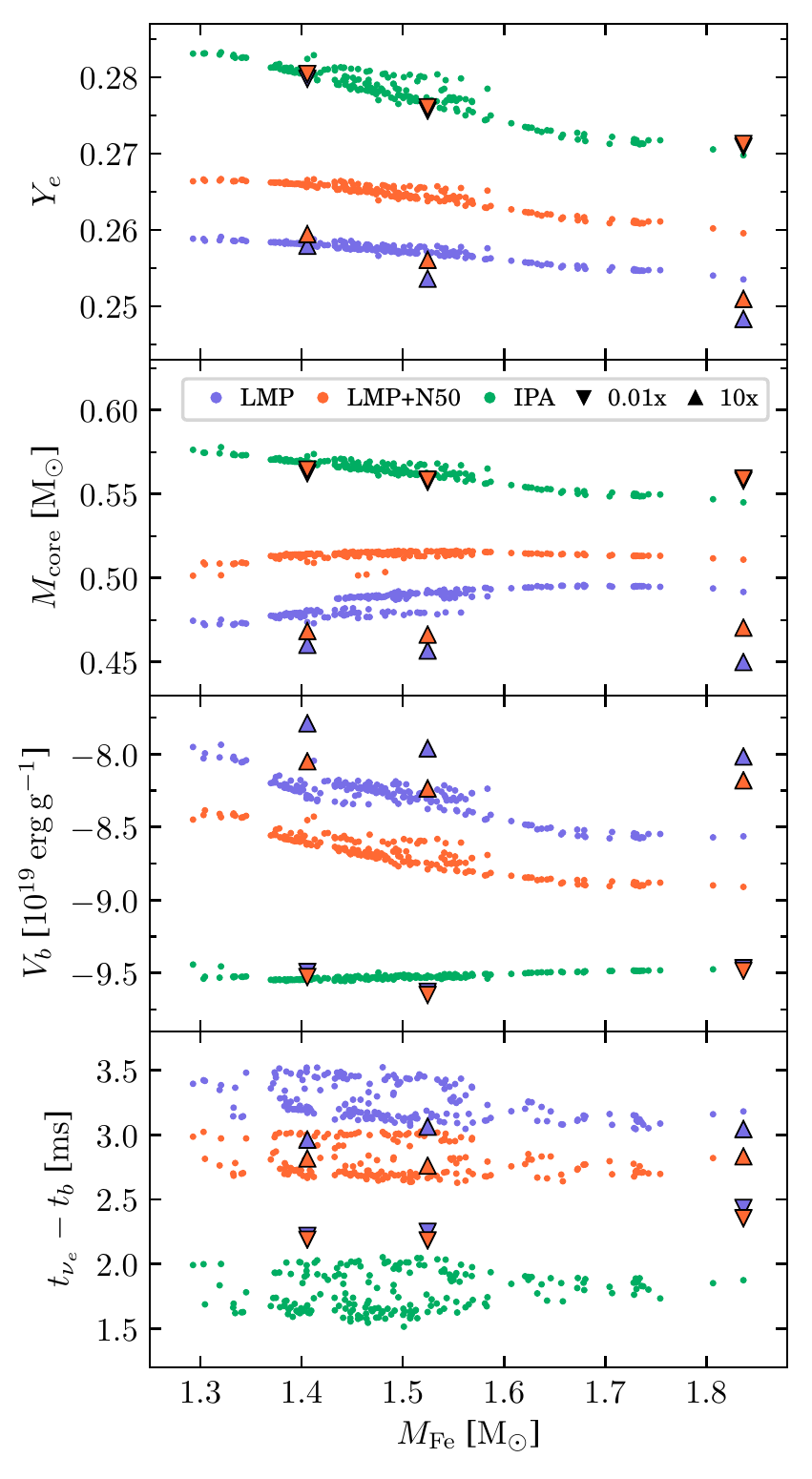}
  \caption{Core bounce properties versus progenitor iron core mass, \Mfe{}, for all simulations (\S~\ref{ss.results.population}).  
  From top to bottom: electron fraction at bounce in the $M=\SI{0.1}{\msun}$ mass shell (\Yebounce{}); inner core mass at bounce (\Mcore{}); gravitational potential at the shock at bounce (\gpot{}); and convergence time of the shock with the \nue{} sphere, relative to bounce (\nushock{}).
  The \onex{} and \tenx{} rate-scaled models are marked by downward and upward-pointing triangles, respectively (appearing from left to right: \SIlist{12;20;40}{\msun}).
  In most cases, the differences between EC rates are larger than the dependence on stellar progenitor.
  }
  \label{fig:pop_stats}
\end{figure}

%++++++++++++++++++++++++++++++++++++++++++++++++++++
%   (figure) Pop delta
%++++++++++++++++++++++++++++++++++++++++++++++++++++
\begin{figure}[t!]
  \centering
  \includegraphics[width=\linewidth]{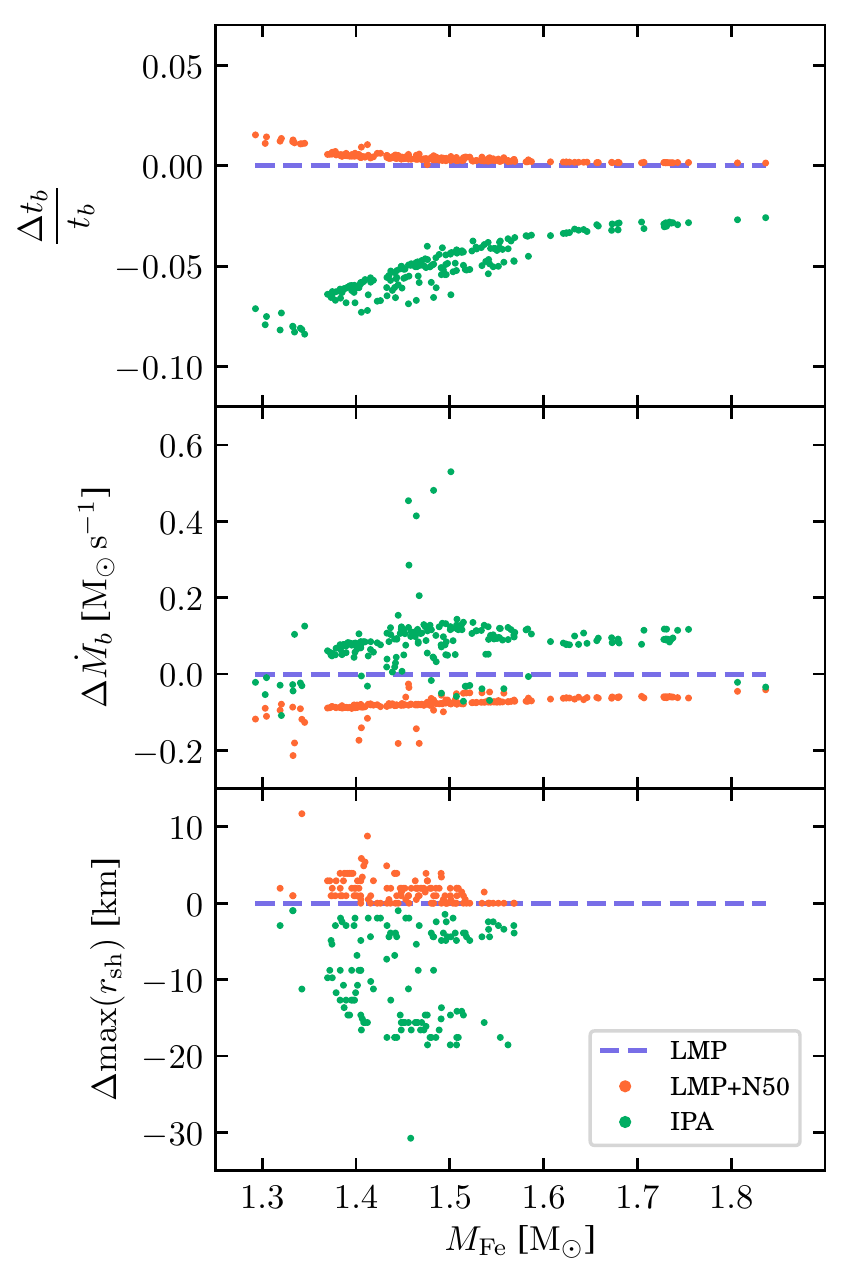}
  \caption{Fractional and absolute differences of LMP+N50 and IPA models relative to LMP, versus progenitor iron core mass} (\S~\ref{ss.results.population}).
  The quantities compared, from top to bottom: time to bounce from start of simulation (\tbounce{}); accretion rate through $r = \SI{500}{km}$ at bounce (\mdotbounce{}); and maximum shock radius reached for failed explosion models (\rshmax{}).
  Note the grid resolution of the simulation limits the precision of \rshock{} here to $\approx \SI{0.5}{km}$.
  \label{fig:pop_delta}
\end{figure}

%++++++++++++++++++++++++++++++++++++++++++++++++++++
%   (figure) Pop remnants
%++++++++++++++++++++++++++++++++++++++++++++++++++++
\begin{figure}[t!]
  \centering
  \includegraphics[width=\linewidth]{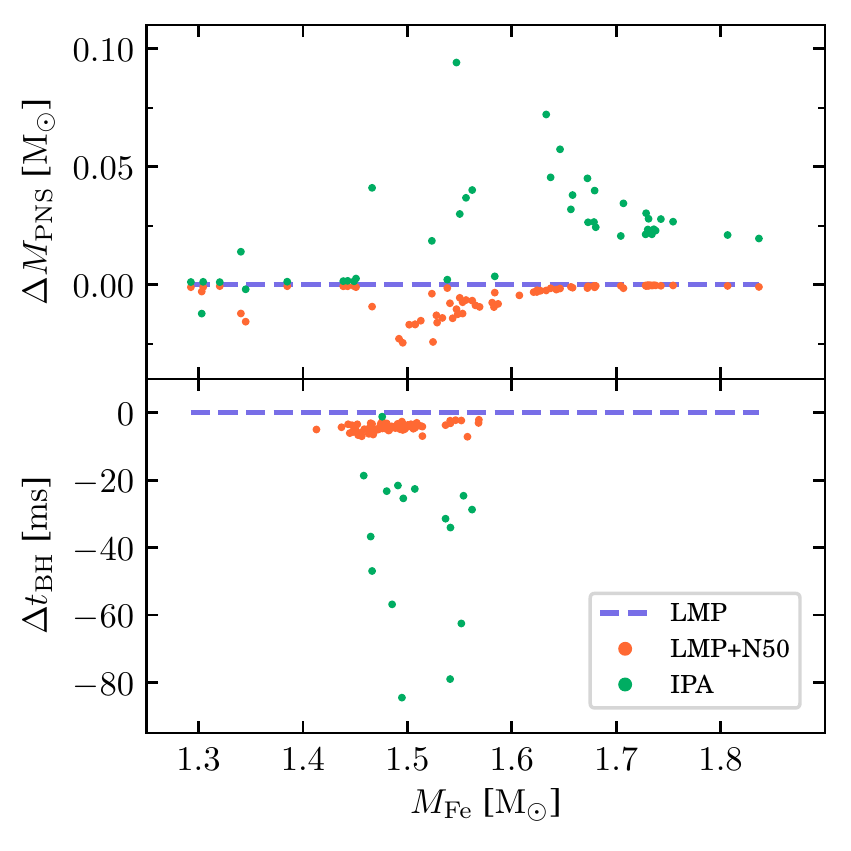}
  \caption{Absolute difference of compact remnant properties relative to LMP (\S~\ref{ss.results.remnants}), versus progenitor iron core mass.  
  Top: proto-neutron star mass at the end of the simulation (\MPNS{}) for exploding models.
  Bottom: time from bounce to BH formation (\tBH{}).
  }
  \label{fig:pop_remnants}
\end{figure}

%==========================================
%         Population Comparisons
%==========================================
\subsection{Population Comparisons}
\label{ss.results.population}
A selection of properties at core bounce for all 200 progenitors are plotted versus the progenitor iron core mass, \Mfe{}, in Figure~\ref{fig:pop_stats}.
These quantities in particular demonstrate large differences between the EC rates compared to differences between the progenitors.

% ye 
The electron fraction of the inner core at bounce, \Yebounce{}, is taken at the $M = \SI{0.1}{\msun}$ enclosed mass coordinate (see also Fig.~\ref{fig:bounce_profiles}).
The IPA models have the largest \Yebounce{} (i.e., weakest deleptonization), followed by LMP+N50 and LMP.

% inner core mass 
The extent of deleptonization translates directly into the inner core mass at bounce, \Mcore{}.
The IPA rates produce systematically larger \Mcore{} than LMP by around \SI{0.08}{\msun} ($\approx \SI{15}{\percent}$), whereas LMP+N50 are around \SI{0.03}{\msun} ($\approx \SI{5}{\percent}$) larger.

% grav potential
The density profile at bounce (Fig.~\ref{fig:bounce_profiles}) determines the gravitational potential at the shock, \gpot{}.
Following previous trends, IPA results in a potential around $\SI{15}{\percent}$ deeper than LMP, compared to LMP+N50, which is consistently $\approx \SI{5}{\percent}$ deeper.

% nue shock
Also shown is \tnue{}, the time when the shock crosses the \nue{} sphere immediately following the bounce.
IPA reaches \tnue{} consistently $\approx \SI{1.5}{ms}$ earlier than LMP, while LMP+N50 is $\approx \SI{0.5}{ms}$ earlier.
This relatively small but persistent difference impacts the neutrino signal of the deleptonization burst (\S~\ref{ss.results.snowglobes}).

% delta values
Quantities that have much larger variation between progenitors than between the EC rates are illustrated in Figure~\ref{fig:pop_delta}.
For clarity, we emphasize the changes due to EC rates by plotting the fractional or absolute difference relative to LMP for each progenitor.
For example, the difference of a given quantity $X$ corresponds to $\Delta X = X - X_\mathrm{LMP}$.
For quantities where the absolute value depends somewhat arbitrarily on the initial conditions, we instead compare the fractional difference, i.e., $\Delta X / X = (X - X_\mathrm{LMP}) / X_\mathrm{LMP}$.

% bounce time
The time of core bounce from the start of the simulation, \tbounce{}, illustrates the speed of collapse from a common starting point.
IPA reaches core bounce between \SIrange{2.5}{8.5}{\percent} earlier than LMP (corresponding to approximately \SIrange{10}{17}{ms}), whereas LMP+N50 reaches bounce $\lesssim \SI{1.5}{\percent}$ later than LMP (corresponding to $\lesssim \SI{2}{ms}$).
In both cases, the difference is the smallest for larger progenitor iron core masses.

% mdot bounce
The mass accretion rate through $r = \SI{500}{km}$ at bounce, \mdotbounce{}, further illustrates the strength of collapse.
Overall, IPA reaches accretion rates around $\SI{0.1}{\msun.s^{-1}}$ larger than LMP, whereas LMP+N50 is $\approx \SI{0.1}{\msun.s^{-1}}$ smaller.

The maximum shock radius reached for failed explosions, \rshmax{}, further supports the differences in shock evolution seen in the reference progenitors (Fig.~\ref{fig:shock_radius}).
IPA reaches the smallest \rshock{}, generally around \SIrange{5}{20}{km} smaller than LMP, whereas LMP+N50 is around \SIrange{0}{5}{km} larger.

%==========================================
%         Population Comparisons
%==========================================
\subsection{Compact Remnants}
\label{ss.results.remnants}
The compact remnant properties are compared in Figure~\ref{fig:pop_remnants} as the absolute difference relative to LMP, as used in Section~\ref{ss.results.population}.

% PNS mass
The PNS mass at the end of the simulation, \MPNS{}, is compared for exploding models.
Here, we define \MPNS{} as the baryonic mass contained in the region above a density of $\SI{1e12}{\density}$.

IPA tends to produce a heavier PNS, with progenitors of $\Mfe \gtrsim \SI{1.45}{\msun}$ around \SIrange{0.02}{0.1}{\msun} larger than LMP.
On the other hand, LMP+N50 tends to produce a similar or slightly lighter PNS, at most up to \SIrange{0.01}{0.02}{\msun} lighter than LMP.
All of the exploding models for $\Mfe \lesssim \SI{1.45}{\msun}$ are within \SI{0.02}{\msun} of LMP.

% BH formation time
The post-bounce time to BH formation, \tBH{}, is compared for the subset of models that reach PNS collapse.
Not all failed-explosion models reach PNS collapse within the time simulated (between \SIrange{1}{5}{s} post-bounce).
For the handful of models that do allow comparison with LMP, IPA reaches BH formation around \SIrange{20}{80}{ms} earlier, whereas LMP+N50 is only around \SIrange{2}{7}{ms} earlier.

Assuming the entire star is accreted, there would be no difference in the final BH mass between the EC rates.

%+++++++++++++++++++++++++++++++
%   (figure) Neutrino Burst
%+++++++++++++++++++++++++++++++
\begin{figure*}[htbp]
  \centering
  \includegraphics[width=\linewidth]{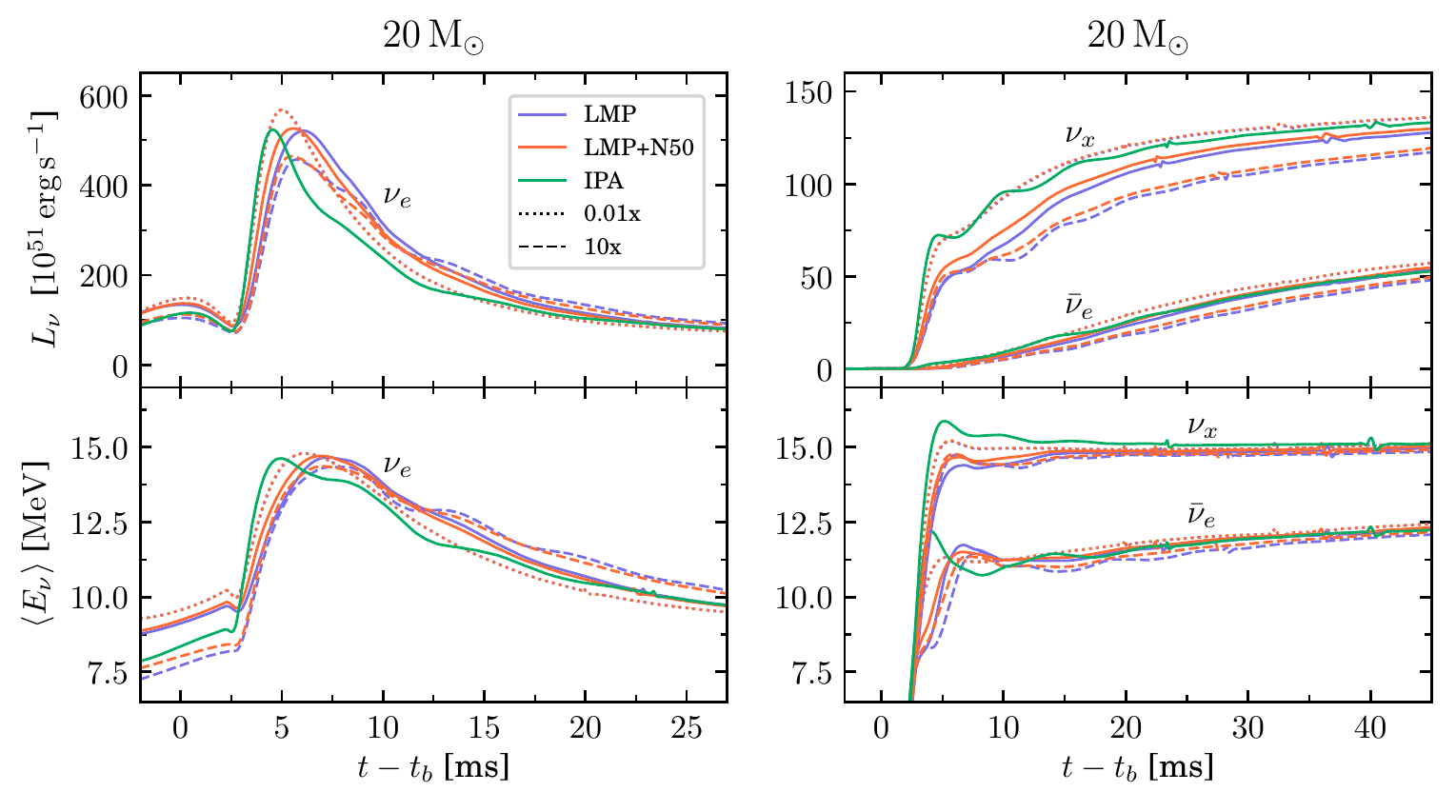}
  \caption{Neutrino emission at $r = \SI{500}{km}$ for the \SI{20}{\msun} progenitor.
  Top row: neutrino luminosity, \Lnu{}. Bottom row: mean neutrino energy, \Enu{}. Left: electron-neutrino (\nue{}) emission. Right: electron antineutrino (\nuebar{}) and heavy-lepton neutrino (\nux{}) emission.  
  Note the different time and \Lnu{} ranges.
  }
  \label{fig:nu_burst}

  \includegraphics[width=\linewidth]{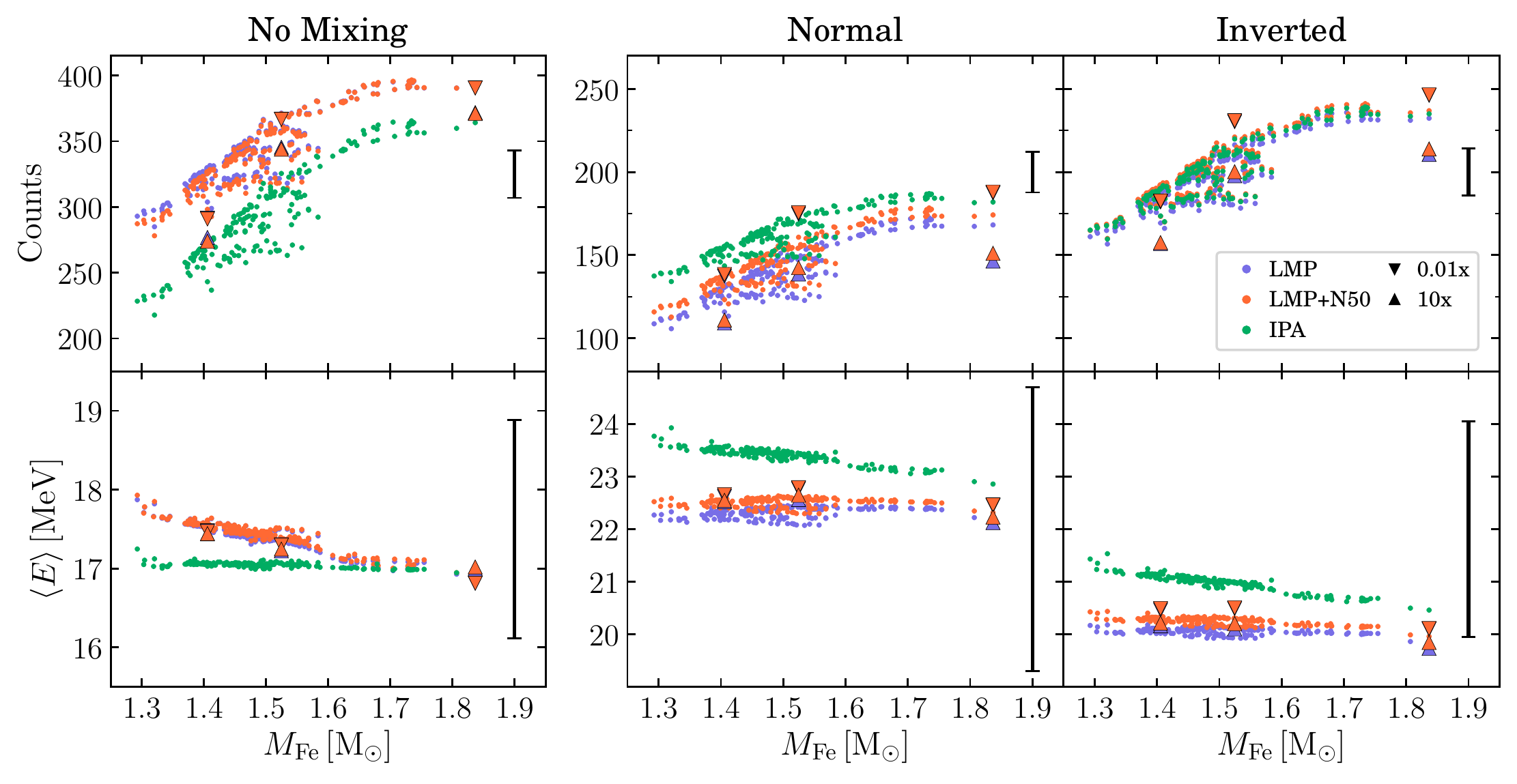}
  \caption{Neutrino burst signal in a DUNE-like liquid argon detector for all 200 progenitors, versus progenitor iron core mass.
  Top row: total neutrino counts.
  Bottom row: mean detected neutrino energy.  
  From left to right are the adiabatic flavor mixing implementations: no flavor mixing, normal mass ordering, and inverted mass ordering. 
  The signal is integrated over \SI{100}{ms} centered on the bounce (\S~\ref{ss.methods.snowglobes}), across all detection channels (Table~\ref{tab:channels}), and assuming a distance of \SI{10}{kpc}.
  The error bars show typical $1\sigma$ uncertainties due to Poisson counting statistics alone.
  Note the different $y$-axis ranges. 
  }
  \label{fig:snowglobes}
\end{figure*}

%==========================================
%         Neutrino Signal
%==========================================
\subsection{Neutrino Signal}
\label{ss.results.snowglobes}
% neutrino lightcurves
The neutrino emission at $r = \SI{500}{km}$ is shown in Figure~\ref{fig:nu_burst} for the \SI{20}{\msun} progenitor.
The three baseline sets reach similar peak electron-neutrino luminosities of $\Lnue \approx \SI{5e53}{\luminosity}$.
The \onex{} luminosities peak roughly \SI{10}{\percent} higher, whereas the \tenx{} peak roughly \SI{10}{\percent} lower.
Additionally, the IPA and \onex{} models peak slightly earlier than the LMP, LMP+N50, and \tenx{} models, which have \nue{} emission spread out to later times.
The mean neutrino energies, \Enu{}, follow a similar pattern.
The \nuebar{} and \nux{} emission is largely reversed, with IPA and \onex{} generally having the largest luminosities and mean energies, followed by LMP+N50, LMP, and \tenx{}.

% snowglobes
The predicted neutrino burst signal in a DUNE-like detector is shown for all 612 models across 200 progenitors in Figure~\ref{fig:snowglobes}.
For all three MSW flavor-mixing cases, there are consistent differences in the detected neutrinos between the EC rate sets.

Overall, LMP and LMP+N50 produce similar total neutrino counts and mean detected energies, \nuEtot{}, with larger differences for IPA, particularly in \nuEtot{}.
When no flavor mixing is assumed, IPA results in the lowest counts and \nuEtot{}.
The effect is reversed when flavor mixing is included for both normal and inverted neutrino mass ordering.
Overall, the inclusion of flavor mixing results in fewer counts and larger \nuEtot{}.

%==============================================================
%         Discussion
%==============================================================
\section{Discussion}
\label{s.discussion}
The choice of EC rates has a clear impact on our simulations of CCSNe.
Perhaps the starkest difference in outcome is whether the models undergo successful shock revival and explosion.
This variation in outcome can largely be traced to the pre-bounce core-collapse phase, where the EC rates control deleptonization and set the conditions at the core bounce for the subsequent shock evolution.
The EC rates also influence the formation of the compact remnant and the observable neutrino signals.

%==========================================
%         Collapse and bounce
%==========================================
\subsection{Collapse and Bounce}
\label{ss.discussion.collapse}
Due to the lack in IPA of forbidden transitions and thermal unblocking effects, the rates are limited to EC on free protons when the average nucleus has a neutron number of $N \geq 40$ \citep{hix:2003}.
Because protons are less abundant than neutron-rich nuclei at high densities, the total number of ECs are suppressed, resulting in a larger \Ye{} in the inner core at bounce compared to LMP and LMP+N50, which do allow EC for $N \geq 40$.
On the other hand, ECs are actually enhanced in IPA at low densities below the $N = 40$ threshold because the rates are overestimated compared to the LMP-based tables \citep{lentz:2012}.

The combined effect is a larger \Ye{} and \Yl{} for IPA in the inner core region ($\rho \gtrsim \SI{1e12}{\density}$) and lower values in the outer core compared to LMP and LMP+N50 (Fig.~\ref{fig:bounce_profiles}~and~\ref{fig:yl_dens}).
This enhanced deleptonization of matter passing through lower densities accelerates the collapse to core bounce, leading to stronger accretion rates, larger infall velocities, larger inner core mass, deeper gravitational potential, and higher densities (Fig.~\ref{fig:pop_stats}~and~\ref{fig:pop_delta}).

These bounce profile differences between the IPA and LMP-based rates are well-established in the literature, noted by \citet{langanke:2003} and \citet{hix:2003}, and reproduced in subsequent studies \citep[e.g.,][]{lentz:2012, sullivan:2016, richers:2017, pascal:2020}.

%==========================================
%         Shock and Explosion
%==========================================
\subsection{Shock Evolution and Explosion Outcome}
\label{ss.discussion.shock-outcome}
Initially, owing to the larger inner core mass, the IPA shock has more kinetic energy and less overlying material to pass through than LMP and LMP+N50.
Competing with these favorable conditions, however, are the higher densities (and thus deeper gravitational potential), faster infall of the overlying material (Fig.~\ref{fig:bounce_profiles}~and~\ref{fig:pop_stats}), and stronger neutrino cooling.
Despite rapid expansion at early times in IPA, the shock is soon overwhelmed by accretion and neutrino cooling, leading to earlier stalling at smaller radii (Fig.~\ref{fig:pop_delta}).
Compounded by the smaller gain region now available for neutrino heating, the result is an earlier shock recession or a delayed explosion (Fig.~\ref{fig:shock_radius}).
By contrast, LMP+N50 tends to have more favorable conditions for a successful explosion, with the slowest collapse to bounce and smallest accretion rates (Fig.~\ref{fig:pop_delta}).

The impact of the EC rates on shock evolution is borne out by the incidence of successful explosions.
Of the 73 progenitors with at least one explosion among the EC rate sets, LMP+N50 explodes in all 73 and LMP explodes in 70.
In contrast, IPA explodes in only 43 of these cases, and for no progenitor is IPA the sole explosion.

%==========================================
%         Compact Remnants
%==========================================
% compact remnants
\subsection{Compact Remnants}
\label{ss.discussion.remnants}
The EC rates also impact the formation of the compact remnant (\S~\ref{ss.results.remnants}).
In successful explosions, the PNS mass, \MPNS{}, is effectively determined by the total mass accreted through the shock before shock revival unbinds the remaining material.
The two deciding factors are thus the accretion rate and the elapsed time before shock runaway.
IPA typically experiences both higher accretion rates (Fig.~\ref{fig:pop_delta}) and later shock revival (Fig.~\ref{fig:shock_radius}), resulting in the largest \MPNS{} (Fig.~\ref{fig:pop_remnants}).
The converse effects on LMP+N50 result in smaller \MPNS{}.

In the case of failed explosions, the conditions at core bounce influence the evolution and eventual collapse of the PNS.
The larger core \Ye{} and \Mcore{} in IPA results in a stronger initial shock, producing a larger PNS radius and stronger \nux{} radiation (Fig.~\ref{fig:nu_burst}).
The subsequent cooling of the PNS results in IPA reaching collapse approximately \SIrange{20}{80}{ms} earlier than LMP (Fig.~\ref{fig:pop_remnants}).
LMP+N50 has only marginally more efficient PNS cooling than LMP, and collapses at most a few milliseconds earlier.
A change in the BH formation time would alter the shutoff of multimessenger signals from neutrinos and gravitational waves.

%==========================================
%         Neutrino Emission
%==========================================
\subsection{Neutrino Emission}
\label{ss.discussion.neutrinos}
The effects of EC on deleptonization and shock formation also manifest in the neutrino emission around bounce (\S~\ref{ss.results.snowglobes}).
For IPA, a larger \Mcore{} at bounce and stronger initial shock, combined with smaller neutrino spheres due to lowered opacities, leads to a faster convergence of the shock with the \nue{} sphere, producing an earlier peak in \Lnue{} (Fig.~\ref{fig:nu_burst}).
The mean time between core bounce and the shock reaching the \nue{} sphere was $1.8 \pm \SI{0.1}{ms}$ for IPA, $3.3 \pm \SI{0.1}{ms}$ for LMP, and $2.8 \pm \SI{0.1}{ms}$ for LMP+N50 ($1\sigma$ standard deviations; Fig.~\ref{fig:pop_stats}).
These differences in \Lnue{} have been noted in previous EC rate studies \citep[e.g.,][]{hix:2003, lentz:2012, sullivan:2016, pascal:2020}.

For LMP and LMP+N50, the extended emission of \nue{} at larger \Lnue{} and \Enue{} results in higher detected neutrino counts and energies in a DUNE-like liquid argon detector when no flavor mixing is assumed (Fig.~\ref{fig:snowglobes}).

When MSW flavor mixing is included with normal neutrino mass ordering, approximately \SI{98}{\percent} of the emitted \nue{} are converted to \nux{}, and vice versa.
This reduces the number of \nue{} available for detection in the dominant \nuecc{} channel (Table~\ref{tab:channels}), resulting in fewer total counts.
There is also a shift to larger \nuEtot{} because most of the \nue{} that are now detected originate as high-energy \nux{}.
Because IPA has larger emitted \Lnux{} and \Enux{}, it now has the highest counts and \nuEtot{}.

The inverted mass ordering case is somewhat intermediate, with only \SI{70}{\percent} of the emitted \nue{} converted to \nux{}, and vice versa.
These favorable survival probabilities result in overall counts and \nuEtot{} that are between the previous two cases.
This appears to roughly coincide with the \textit{crossover} point, where all three EC rates produce very similar counts.

The large error bars from Poisson statistics alone suggest that these differences would be difficult to detect under these narrow assumptions, especially for \nuEtot{}, even if the progenitor was known.
Additionally, there are degeneracies between the EC rate and the impacts of the progenitor star, neutrino mass hierarchy, and potentially the nuclear EOS (not investigated here).
Reducing the uncertainties via larger count numbers could be achieved by combining measurements from multiple neutrino detectors, or if the supernova occurred closer than the assumed \SI{10}{kpc}.
The degenerate signals might be broken by incorporating into the analysis: additional parts of the neutrino lightcurve \mbox{\citep[e.g.,][]{segerlund:2021}}; additional detectors sensitive to other neutrino flavors (e.g., water Cherenkov detectors); or, given sufficient counts, the full neutrino spectrum instead of an average energy.

%==============================================================
%         Conclusion
%==============================================================
\section{Conclusion}
\label{s.conclusion}
We have produced a suite of 612 one-dimensional CCSN simulations, using three sets of EC rates and 200 stellar progenitors between \SIrange{9}{120}{\msun}.

% EC tables
The three EC rate sets were (\S~\ref{ss.methods.ecrates}): an IPA \citep{bruenn:1985}; a microphysical library with parametrized rates in the high-sensitivity $N = 50$ region \citep[LMP;][]{sullivan:2016,titus:2018}; and the same library with updated $N = 50$ rates \citep[LMP+N50;][]{titus:2019}.

% overview of results
Of the 200 progenitors, there were 43 successful explosions for the IPA set, 70 for LMP, and 73 for LMP+N50.
In general, the IPA models reached smaller shock radii and exploded later than their LMP and LMP+N50 counterparts (\S~\ref{ss.results.reference}).
Of the latter two, LMP+N50 appeared marginally more favorable to explosion, with larger shock radii and earlier shock runaway than LMP.

% collapse and bounce
At core bounce, IPA typically had the largest inner core mass, electron fraction, density, accretion rate, infall velocity, and gravitational potential (Fig.~\ref{fig:bounce_profiles},~\ref{fig:pop_stats},~and~\ref{fig:pop_delta}).
The next largest values were generally LMP+N50 followed by LMP, although LMP+N50 changed places with LMP for the collapse time and accretion rate (Fig.~\ref{fig:pop_delta}).
The standard ordering was also reversed for \Ye{} in the outer core, where IPA had the lowest values and LMP the highest (Fig.~\ref{fig:yl_dens}).

% compact remnants
For exploding progenitors, IPA produced a PNS mass around $\SIrange{0.02}{0.1}{\msun}$ larger than LMP due to higher accretion rates and delayed shock revival, and LMP+N50 was typically $\lesssim \SI{0.02}{\msun}$ smaller (Fig.~\ref{fig:pop_remnants}).
For failed explosions, enhanced PNS cooling in IPA resulted in a collapse to BH roughly \SIrange{20}{80}{ms} earlier than LMP, whereas LMP+N50 was at most a few milliseconds earlier.

% neutrinos
Without flavor mixing, the extended \nue{} emission of LMP and LMP+N50 following bounce (Fig.~\ref{fig:nu_burst}) resulted in higher detected counts and mean energies than IPA in a DUNE-like liquid argon detector (Fig.~\ref{fig:snowglobes}).
Conversely, when adiabatic flavor mixing is included, the enhanced \nux{} emission in IPA  is converted to \nue{}, resulting in higher counts and energies than LMP and LMP+N50.
Given only $\sim 10^2$ counts, however, these differences were typically smaller than the estimated uncertainties.

% interpretation
All of these results largely stem from the total rate of electron captures at different densities during collapse (\S~\ref{ss.discussion.collapse}).
For IPA, the total EC rate is overestimated at lower densities, but subsequently underestimated at higher densities due to Pauli blocking on a mean nucleus of $N \geq 40$.
The LMP-based rates unblock EC for neutron-rich nuclei, and so deleptonization proceeds further than IPA during collapse.
The updated $N = 50$ rates in LMP+N50 are lower than the parametrized rates in LMP, producing an intermediate case between LMP and IPA (Fig.~\ref{fig:pop_stats}).

% limitations
It is important to emphasize the limitations of our study.
While our \stir{} framework \citep{couch:2020} approximates the effects of turbulence in 1D, there is ultimately no substitute for multidimensional simulations.
Only high-fidelity 3D simulations can hope to fully capture the interplay between fluid instabilities, magnetohydrodynamics, and neutrino transport in CCSNe \citep[e.g.,][]{hanke:2013, lentz:2015, oconnor:2018b, summa:2018, muller:2019}.
Our progenitors from \citet{sukhbold:2016} were 1D, solar-metallicity, nonrotating, and nonmagnetic, and do not represent the full variety of stellar populations.
The progenitor models also used microphysical EC rates, and so the sudden transition to approximate rates in our IPA models is somewhat artificial.
Although we accounted for adiabatic flavor mixing when calculating neutrino signals, there remains copious uncertainty around the effects of flavor oscillations, which were not included in our simulations.

% EOS
We note that our study only considered the SFHo nuclear EOS, as previous works have demonstrated that the EOS dependence of the collapse and early post-bounce phase ($\lesssim \SI{100}{ms}$) is dwarfed by the impact of EC rate uncertainties \mbox{\citep{sullivan:2016}}.
We posit that the qualitative differences seen here between the EC rates beyond $\approx \SI{100}{ms}$ post-bounce are unlikely to be dramatically altered by the EOS.
Nevertheless, the choice of EOS can alter the nuclear abundances that EC acts upon \mbox{\citep{nagakura:2019}}, and it would be valuable to test our assumption in a future study.

% New Giraud rates
Finally, the updated rates in LMP+N50 do not include temperature dependence effects \citep{dzhioev:2020}.
Very recently, \citet{giraud:2022} reported new finite-temperature calculations for the $N=50$ region, with rates around an order of magnitude higher than LMP+N50 and about a factor of 5 below the LMP rates. Their simulations suggest that CCSN properties would be intermediate between the LMP and LMP+N50 models presented here, which are already in relatively good agreement compared to the commonly used IPA rates. Work is underway to incorporate these new rates into \nulib{} so that they can be freely used in future CCSN simulations.

% closing statement
EC plays a central role in deleptonization, shock formation, and neutrino production during core-collapse.
Our study has explored the effects of updated EC rates in the high-sensitivity $N = 50$ region, including a detailed comparison between microphysical rates and a simple IPA.
By producing simulations across 200 progenitors, we have shown there are clear, systematic impacts of EC rates on the core structure, shock dynamics, and neutrino signals throughout the CCSN mechanism.

%==================================
%         End matter
%==================================
\begin{acknowledgments}
This work was supported in part by Michigan State University through computational resources provided by the Institute for Cyber-Enabled Research.
SMC is supported by the U.S. Department of Energy, Office of Science, Office of Nuclear Physics, under Award Numbers DE-SC0015904 and DE-SC0017955.
RT and RZ are supported by the US National Science Foundation under Grants PHY-1913554 (Windows on the Universe: Nuclear Astrophysics at the NSCL), PHY-1430152 (JINA Center for the Evolution of the Elements), and PHY-1927130 (AccelNet-WOU: International Research Network for Nuclear Astrophysics [IReNA]).
EOC is supported by the Swedish Research Council (Project No. 2020-00452).
\end{acknowledgments}

\software{
  \flash{}\footnote{\url{https://flash.rochester.edu/site/}} \citep{fryxell:2000,dubey:2009}, 
  NSCL Weak-rate Library v1.2 \citep{sullivan:2015},
  \nulib{} \citep{oconnor:2015},
  \snowglobes{} \citep{scholberg:2012},
  \textsc{GLoBES} \citep{huber:2005},
  \textsc{Matplotlib}\footnote{\url{https://matplotlib.org}} \citep{hunter:2007}, 
  \textsc{NumPy}\footnote{\url{https://www.numpy.org}} \citep{harris:2020}, 
  \textsc{SciPy}\footnote{\url{https://www.scipy.org}} \citep{virtanen:2020},
  \textsc{yt}\footnote{\url{https://yt-project.org}} \citep{turk:2011},
  \textsc{Pandas}\footnote{\url{https://pandas.pydata.org}} \citep{pandas:2022},
  \textsc{xarray}\footnote{\url{https://xarray.pydata.org}} \citep{hoyer:2017,hoyer:2020},
  \textsc{Astropy}\footnote{\url{https://www.astropy.org}} \citep{astropy:2013, astropy:2018},
  \flashbang{}\footnote{\url{https://github.com/zacjohnston/flashbang}} \citep{johnston:2022a},
  \flashsnowglobes{}\footnote{\url{https://github.com/zacjohnston/flash_snowglobes}} \citep{johnston:2022b}.
  }

\bibliography{main}

%==============================================================
%         Appendix
%==============================================================
\appendix

%==================================
%         Flavor Mixing
%==================================
\section{Neutrino Flavor Mixing}
\label{a.mixing}
When calculating detectable neutrino counts with \snowglobes{} (\S~\ref{ss.methods.snowglobes}), we account for adiabatic neutrino flavor conversions due to MSW matter effects \citep{dighe:2000}. 
Following similar approaches in \citet{nagakura:2021} and \citet{segerlund:2021}, we calculate the neutrino flux at Earth, $F_i$, for each neutrino flavor $i$, using
\begin{align}
    \Fe &= p \Fe^0 + (1 - p) \Fx^0, \\
    \Febar &= \bar{p} \Febar^0 + (1 - \bar{p}) \Fxbar^0, \\
    \Fx &= \frac{1}{2} (1 - p) \Fe^0 + \frac{1}{2} (1 + p) \Fx^0, \\
    \Fxbar &= \frac{1}{2} (1 - \bar{p}) \Febar^0 + \frac{1}{2} (1 + \bar{p}) \Fxbar^0,
\end{align}
where $p$ and $\bar{p}$ are the survival probabilities, and $F^0_i$ are the emitted fluxes for each flavor $i$.
Under normal neutrino mass ordering, the survival probabilities are given by
\begin{align}
  p &= \sin^2 \theta_{13} \approx 0.02, \\
  \bar{p} &= \cos^2 \theta_{12} \cos^2 \theta_{13} \approx 0.69,
\end{align}
and under inverted mass ordering by
\begin{align}
  p &= \sin^2 \theta_{12} \cos^2 \theta_{13} \approx 0.29 \\
  \bar{p} &= \sin^2 \theta_{13} \approx 0.02,
\end{align}
where we use mixing parameters of $\sin^2 \theta_{12} = 0.297$ and $\sin^2 \theta_{13} = 0.0215$ \citep{capozzi:2017}. The no-mixing case is equivalent to $p = \bar{p} = 1$.
Note that our simulations use the combined heavy-lepton species $\nux = \{ \nu_\mu$, $\nu_\tau$, $\bar{\nu}_\mu$, $\bar{\nu}_\tau \}$, and thus we assume 
\begin{equation}
    \Fx^0 = \Fxbar^0 \propto \frac{1}{4} \Lnux,
\end{equation}
where $\Lnux$ is the heavy-lepton neutrino luminosity from the simulation (Fig.~\ref{fig:nu_burst}).
The separated heavy-flavor inputs to \snowglobes{} are then 
\begin{align}
F_\mu = F_\tau &= \Fx, \\ 
\bar{F}_\mu = \bar{F}_\tau &= \Fxbar.
\end{align}

The code used for these calculations is publicly available as a \textsc{Python} package, \flashsnowglobes{} \citep{johnston:2022b}.

%==================================
%         Data Availability
%==================================
\section{Data Availability}
\label{a.data}
The data for all simulations presented in this work are publicly available in a Mendeley Data repository \citep{johnston:2022c}.
The dataset includes summarized simulation results, radial profiles at core bounce, time-dependent quantities (e.g., shock radius and neutrino luminosities), and time-binned \snowglobes{} neutrino counts and energies.

Much of the model analysis and plotting were performed using our publicly available \textsc{Python} package \flashbang{} \citep{johnston:2022a}.

\end{document}